\documentclass[10pt,journal,twoside,twocolumn]{IEEEtran}

\usepackage{amsmath,amsfonts}
\usepackage{textcomp}
\usepackage{stfloats}
\usepackage{verbatim}
\usepackage{graphicx}
\def\BibTeX{{\rm B\kern-.05em{\sc i\kern-.025em b}\kern-.08em
		T\kern-.1667em\lower.7ex\hbox{E}\kern-.125emX}}
\usepackage{balance}
\usepackage{bm}
\usepackage{subfigure}
\usepackage[hidelinks,
colorlinks=true,
allcolors=black,
pdfstartview=Fit,
breaklinks=true]{hyperref}
\usepackage{cite}
\allowdisplaybreaks[4]

\begin{document}
	\title{Optimization of RIS Configurations for Multiple-RIS-Aided mmWave Positioning Systems based on CRLB Analysis}
	\author{Yu Liu, Sheng Hong, Cunhua Pan, Yinlu Wang, Yijin Pan, and Ming Chen
	\thanks{\textit{(Corresponding author: Cunhua Pan)}}
	\thanks{C. Pan is with the School of Electronic Engineering and Computer Science at Queen Mary University of London, London E1 4NS, U.K. (e-mail: c.pan@qmul.ac.uk).}
	\thanks{Y. Liu, Y. Wang, Y. Pan and M. Chen are with the National Mobile Communications Research Laboratory, Southeast University, Nanjing 210096, China. (e-mail:\{liuyu\_1994, yinluwang, chenming, panyj\}@seu.edu.cn).}
	\thanks{S. Hong is with School of Information Engineering at Nanchang University, Nanchang 330100, China. (e-mail: shenghong@ncu.edu.cn)}
	}
	\maketitle

\begin{abstract}
Reconfigurable intelligent surface (RIS) is a promising technology for future millimeter-wave (mmWave) communication systems. However, its potential benefits of adopting RIS for high-precision positioning in mmWave systems are still less understood. In this paper, we study a multiple-RIS-aided mmWave positioning system and derive the Cram$\rm{\acute{e}}$r-Rao error bound. Based on the derived bound, we optimize the phase shift of the RISs by the particle swarm optimization (PSO) algorithm.  Numerical results have demonstrated the advantages of using multiple RISs in enhancing the positioning accuracy in mmWave systems. 
\end{abstract}


\section{Introduction}
In the current 4G/5G wireless networks, local positioning system (LPS) has attracted extensive research interests due to its key role in providing accurate beamforming \cite{localization_survey}. Recent research efforts in LPS have successfully reduced the positioning error to the centimeter level with a single base station in mmWave communications \cite{shahmansoori2017position}. However, this level of accuracy cannot meet the requirements demanded by emerging applications in future B5G/6G networks, such as smart factories, autonomous driving, and augmented reality. In these scenarios,  the wireless propagation links suffer from severe propagation loss and obstacles blockage when operating in mmWave and Terahertz frequency band, which will deteriorate the positioning accuracy and reliability \cite{mmwavechannel, abu2018error}. Therefore, it is imperative to improve the positioning accuracy and reliability of LPS for the next-generation wireless networks.

Fortunately, the promising reconfigurable intelligent surface (RIS) technique could provide a stable and high-precision positioning performance in a cost-effective and energy-efficient manner. An RIS is a planar surface composed of a large number of low-cost and passive reflecting elements, which can adjust the amplitudes and phase shifts of the incident signals \cite{CPan,survey2,near_field,indoor_zhang}. Hence, RIS can be deployed to create an alternative link to bypass the blockage of the direct link, which enables the LPS to recover positioning functionality in dead zones. Unlike fixed scatters in the environment, the RIS could be envisioned as a controllable scatter by adjusting the phase shifts of the reflection elements. Furthermore, an RIS with large aperture provides a higher beamforming gain as well as a higher angle resolution, which is appealing for positioning. It is expected that RIS could improve the positioning accuracy, which is measured by the Cram$\rm{\acute{e}}$r-Rao lower bound (CRLB). 

The authors in \cite{JH2020} investigated the advantages of deploying a single RIS with a uniform linear array (ULA) for a two-dimensional positioning system. This work was extended to a three-dimensional scenario with a planar RIS in \cite{2021RISwangrui}. However, to the best of our knowledge, there are no existing works that adopted multiple RISs with varying sizes for positioning in mmWave systems. In addition, the existing works designed the phase shifts of the RIS using the beam matching method, which is a suboptimal choice in terms of minimizing the CRLB. 

The contributions of this paper are summarized as follows:  
\begin{enumerate}
\item We derive the position and orientation error bounds from the CRLB of a multiple-RIS-aided positioning system.  
\item Due to the complicated expression of CRLB, we implement the heuristic algorithm like particles swarm optimization (PSO) algorithm to achieve the nearly optimal solution.
\item The impact of various system parameters, such as the number of the RISs and the size of the RIS, on the localization error bounds is studied in the simulations. 
\end{enumerate}

\section{System Model}
\begin{figure}[htbp]
	\centering
	\includegraphics[width=0.8 \linewidth]{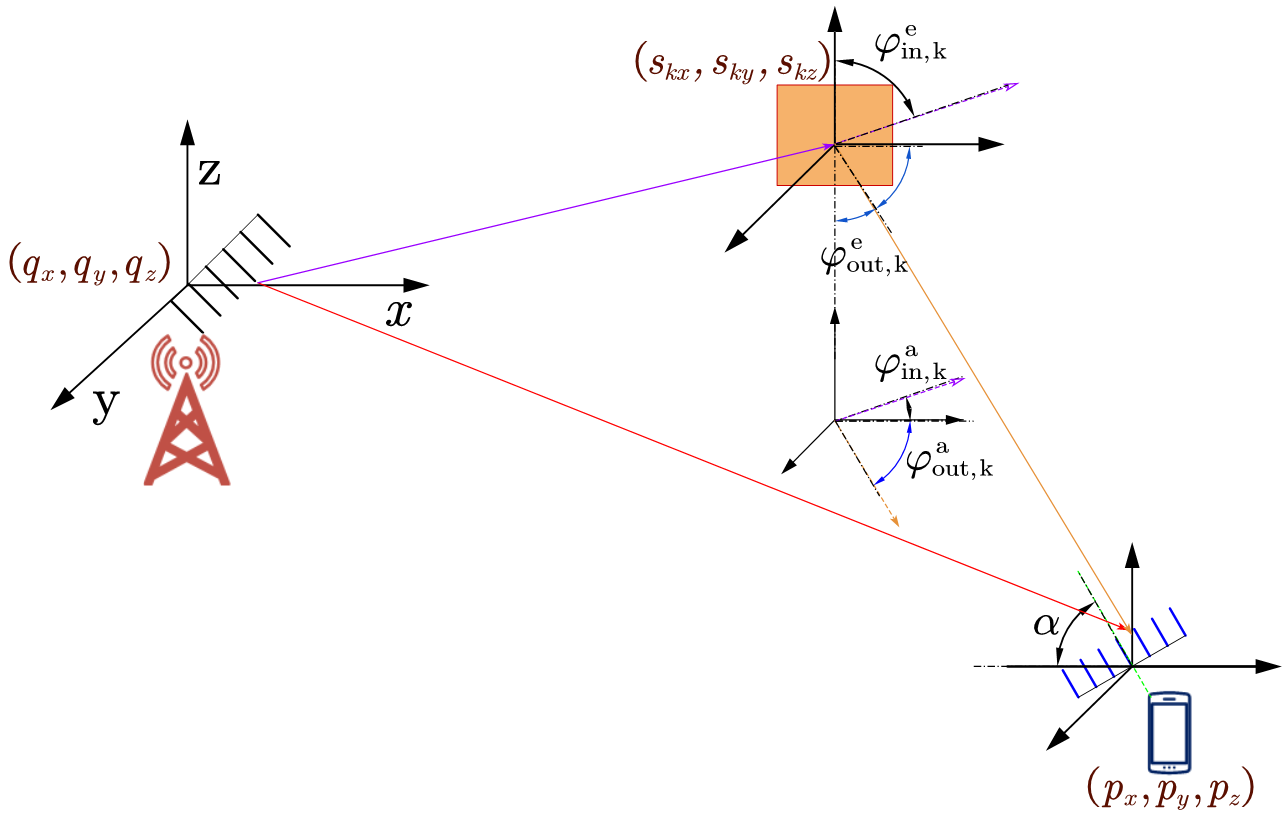}
	\caption{The positioning  simulation scenario with three RISs}
	\label{img-1d3r}
\end{figure}

We consider a multiple-RIS-aided 3D positioning system shown in Fig. \ref{img-1d3r}, which consists of a base station (BS) equipped with a ULA of $N_t$ antennas,  a mobile user (MU) with a ULA of $N_r$ antennas and $K$ RISs with a uniform planar array (UPA) of $L^2$ reflecting elements. The locations of the BS antenna array, the $k$-th RIS, and the MU are respectively denoted by $\mathbf{q}=\left[q_x,q_y,q_z\right]^T \in R^3$, $\mathbf{s}_k=\left[s_{kx}, s_{ky}, s_{kz}\right]^T \in R^3$ and $\mathbf{p}=\left[p_x,p_y,0\right]^T \in R^3$, where $q_k$ and $s_{kz}$ symbolizes the heights of BS and the $k$-th RIS relative to the MU on the ground. The rotation angle of the MU's antenna array is denoted by $\alpha \in \left(0,\pi\right]$, and no rotation is assumed for BS and RIS arrays. The locations of the BS and the RISs are assumed to be known. Thus, the objective of the positioning system is to estimate the coordinates and rotation angle of the MU.

\subsection{Transmitter Model}
We consider a system operating at a carrier frequency $f_c$ with wavelength $\lambda$. The total bandwidth is $B$ with $N$ orthogonal frequency division multiplexing (OFDM) subcarriers. The transmit signal on the $n$-th subcarrier is denoted by $\mathbf{x}[n]=[x_1[n],\cdots,x_{M_t}[n]]^T$, where $M_t \ll N_t$ denotes the number of beams. To simplify the analysis, it is assumed that the pilot signal has unit power and flat spectrum. Thus, its power spectrum is a constant of $\left|X(\omega)\right|^2=\frac{1}{2\pi B}$ within $\omega \in \left[0,2\pi B \right]$. The transmitted signals over subcarrier $n$ can be expressed as $\bm{F}\mathbf{x}[n]$, where $\mathbf{F}$ is the beamforming matrix defined as $ \mathbf{F} = \left[\mathbf{f}_1,\mathbf{f}_2,\cdots,\mathbf{f}_{M_t}\right] \in C^{N_t \times M_t}$, with $\mathbf{f}_i \in C^{N_t}$ denoting the unit-norm transmitting vector.

\subsection{Wireless Channel Model}
The channel state information (CSI) matrix $\mathbf{H}$ between the BS and the MU could be expressed as the combination of the line-of-sight (LoS) component and the reflecting component, which is given by
\begin{equation}
	\bm{H}[n]=\underbrace{\bm{H}_{0}[n]}_{\text{LoS channel}}+\underbrace{\sum_{k=1}^{K}\bm{H}_{k}[n]}_{\text{reflecting channel}} .
\end{equation}
The expressions of $\bm{H}_{0}[n]$ and $\bm{H}_{k}[n]$ are respectively given by
\begin{align}
	&
	\bm{H}_{0}[n]\!=\!\gamma_0 h_0\bm{\alpha}_{RX}\left(\theta_{RX,0}\right)\bm{\alpha}_{TX}^{H}\left(\theta_{TX,0}\right) e^{j2\pi B \frac{n}{N}\tau_{0}} ,\\
	&\bm{H}_{k}[n]\!=\!\gamma_k h_k\bm{H}_{IM,k}\bm{\Theta}_k \bm{H}_{BI,k} e^{j2\pi B \frac{n}{N}\tau_{k}} , \label{H_RIS,k} 
\end{align}
where $\gamma_k=\sqrt{N_t N_r/\rho_k}$ with $\rho_k$ being the path loss of the $k$-th reflecting channel,  $\tau_k$ is the propagation delay, and $h_k$ is the complex channel gain. In Eq. (\ref{H_RIS,k}),  $\bm{H}_{BI,k}$ is the channel matrix from the BS to the $k$-th RIS, and $\bm{H}_{IM,k}$ is the channel matrix from the $k$-th RIS to the MU, and $\bm{\Theta}_k$ is the diagonal phase shift matrix of the $k$-th RIS. They are respectively given by
\begin{align}
	&\bm{H}_{BI,k}=\bm{\alpha} _{RIS,IN}\left( \varphi _{in,k}^{a},\varphi_{in,k}^{e} \right) \bm{\alpha} _{TX}^{H}\left( \theta _{TX,k} \right) ,  \label{H_BI,k}\\
	&\bm{H}_{IM,k}=\bm{\alpha }_{RX}\left( \theta _{RX,k} \right) \bm{\alpha}_{RIS,OUT}^{H}\left( \varphi _{out,k}^{a},\varphi_{out,k}^{e} \right) , \label{H_IM,k}\\
	&\bm{\Theta}_k=\delta\times diag(e^{j\theta_1},e^{j\theta_2},\cdots,e^{j\theta_{L^2}}) \in C^{L^2\times L^2}. \label{Theta_k}
\end{align}
In Eq. \eqref{H_BI,k} and Eq. \eqref{H_IM,k}, $\bm{\alpha}_{TX}\left(\theta_{TX,k}\right) \in \mathbb{C}^{N_t}$ and $\bm{\alpha}_{RX}\left(\theta_{RX,k}\right) \in \mathbb{C}^{N_r}$ are the antenna response vectors of the transmitter and the receiver respectively, with $\theta_{TX,k}$ denoting the angle of departure (AOD) and $\theta_{RX,k}$ the angle of arrival (AOA). The $i$-th entry of $\bm{\alpha}_{TX}\left(\theta_{TX}\right)$ is 
$ \left[ \bm{\alpha}_{TX}\left(\theta_{TX}\right) \right]_{i}=e^{j(i-1)\frac{2\pi}{\lambda}d\sin(\theta_{TX}) }$
and the $i$-th entry of $\bm{\alpha}_{RX}\left(\theta_{RX}\right)$ is 
$ \left[ \bm{\alpha}_{RX}\left(\theta_{RX}\right)  \right]_{i}=e^{j(i-1)\frac{2\pi}{\lambda}d\sin(\theta_{RX}) } $
, where $d$ is the distance between adjacent antennas and $\lambda=c/f_c$ with $c$ being the speed of light. Also in Eq. \eqref{H_BI,k} and Eq. \eqref{H_IM,k}, $\bm{\alpha} _{RIS,IN}( \varphi _{in,k}^{a},\varphi_{in,k}^{e} )$ and $\bm{\alpha}_{RIS,OUT}( \varphi _{out,k}^{a},\varphi_{out,k}^{e} )$ are the array response vectors of the BS-RIS$_k$ and RIS$_k$-MU links respectively, where $\varphi_{in,k}^{a}$ and $\varphi_{in,k}^{e}$ are respectively the azimuth AOA and elevation AOA at the BS-RIS$_k$ link, $\varphi_{out,k}^{a}$ and $\varphi_{out,k}^{e}$ are respectively the azimuth AOD and elevation AOD at the RIS$_k$-MU link. The $\left[a+(b-1)L\right]$-th elements in $\bm{\alpha} _{RIS,IN}$ and $\bm{\alpha}_{RIS,OUT}$ are respectively given by 
\begin{align}
	\begin{aligned}
		\left[\bm{\alpha}_{RIS,IN}\right]_{a+(b-1)L}=&e^{j\frac{2\pi}{\lambda}(b-1)d\cos\left(\varphi_{in}^{e}\right)}\\
		&\times e^{j\frac{2\pi}{\lambda}d(a-1)\sin\left(\varphi_{in}^{e}\right)\sin\left(\varphi_{in}^{a}\right)} , \\
	\end{aligned} \\
	\begin{aligned}
		\left[\bm{\alpha}_{RIS,OUT}\right]_{a+(b-1)L}=&e^{j\frac{2\pi}{\lambda}(b-1)d\cos\left(\varphi_{out}^{e}\right)}\\
		&\times e^{j\frac{2\pi}{\lambda}d(a-1)\sin\left(\varphi_{out}^{e}\right) \sin\left(\varphi_{out}^{a}\right)} .
	\end{aligned}
\end{align}
In Eq. \eqref{Theta_k}, $\delta=1$ represents the amplitudes of reflecting coefficients and $\theta_i$ is the phase shift of the $i$-th reflecting element.
Based on \cite{cuitiejun2021}, the path loss $\rho_0$ of the LoS channel is modeled as 
\begin{equation}
	\begin{aligned}
		&PL_{LOS}\left(d_0\right)=10\log_{10}\rho_0 \\
		&=10\log_{10}(64\pi^3)+10\alpha_0 \log_{10}d_0+20\log_{10}f_c+\xi_0 ,
	\end{aligned} 
\end{equation}
where $d_0$ is the distance between the MU and the BS, and $\alpha_0$ is the path loss exponent, $\xi_0\sim\mathcal{CN}\left(0,\sigma_{SF,0}^2\right)$ is the log-normal shadow fading, with $\sigma_{SF,0}^2$ denoting the shadowing variance. The path loss $\rho_k$ of the $k$-th RIS-aided link is modeled as 
\begin{equation}
	\begin{aligned}
		&PL_{RIS}\left(d_{k1},d_{k2}\right)=10\log_{10}\rho_k \\
		&=10\log_{10}(64\pi^3)+10\alpha_k \log_{10}(d_{k1}\cdot d_{k2})+40\log_{10}f_c+\xi_k ,
	\end{aligned}
\end{equation}
where $d_{k1}$ is the distance of the BS-RIS link, $d_{k2}$ is the distance of the RIS-MU link, $\alpha_1$ is the path loss exponent of the $k$-th reflecting link, and $\xi_k\sim\mathcal{CN}\left(0,\sigma_{SF,k}^2\right)$ is the log-normal shadow fading, with $\sigma_{SF,k}^2$ denoting the shadowing variance.

\subsection{Receiver Model}
The received signal at the MU is given by 
\begin{equation}
	\bm{y}\left[n\right]=\sqrt{P_{TX}}\bm{H}[n]\bm{F{\rm x}}[n]+\bm{{\rm n}}[n] , \label{receiver}
\end{equation}
where $P_{TX}$ is the transmitter power and $\bm{{\rm n}}\sim \mathcal{N}(0,N_0 \bm{I}_{N_r})$ is the additive white Gaussian noise. In the following, we first derive the CRLB of the position and orientation of the MU based on the received signal model in Eq. (\ref{receiver}). Then, we provide a low-complexity algorithm to optimize the phase shift of the RIS to minimize the CRLB. 

\section{Problem Formulation}
We employ the two-stage approach to estimate the position and orientation of the MU. In the first step, we estimate the channel parameters $\bm{\eta}$ in a general form as 
\begin{equation}
	\hat{\bm{\eta}}=\bm{\eta}+\bm{w} ,
\end{equation}
where $\bm{w} \sim \mathcal{CN}(\bm{H}[n]\bm{F{\rm x}}[n],\bm{\Sigma})$ denotes the estimation error with $\bm{\Sigma}=N_0\bm{I}_{N_r}$ and $\bm{\eta}$ denotes the channel parameters, such as TOA, AOA, AOD and complex channel gain.  Specifically, $\bm{\eta}$ is given by
\begin{equation}
	\bm{\eta}=\left[\bm{\tau},\theta_{TX,0},\bm{\theta}_{RX},\bm{\varphi}_{out}^{a},\bm{\varphi}_{out}^{e}, \bm{h}_R, \bm{h}_I \right]^T ,
\end{equation}
where 
\begin{align}
	&\bm{\tau}=\left[\tau_0, \tau_1,\cdots,\tau_{K}\right]^T  , \\
	&\bm{\theta}_{RX}=\left[\theta_{RX,0},\theta_{RX,1},\cdots,\theta_{RX,K}\right]^T  , \\
	&\bm{\varphi}_{out}^{a}=\left[\varphi_{out,1}^{a},\varphi_{out,2}^{a},\cdots,\varphi_{out,K}^{a}\right]^T  , \\
	&\bm{\varphi}_{out}^{e}=\left[\varphi_{out,1}^{e},\varphi_{out,2}^{e},\cdots,\varphi_{out,K}^{e}\right]^T  , \\
	&\bm{h}_R=\left[h_{R,0}, h_{R,1},\cdots,h_{R,K}\right]^T , \\
	&\bm{h}_I=\left[h_{I,0}, h_{I,1},\cdots,h_{I,K}\right]^T .
\end{align}
In the second step, we estimate the MU's position $\hat{\bm{p}}$ and orientation $\hat{\alpha}$ as follows
\begin{equation}
	\begin{aligned}
		\left[\hat{\bm{p}},\hat{\alpha}\right]&=\mathop{\arg\max}_{\left[\bm{p},\alpha\right]} \quad p\left(\hat{\bm{\eta}}| \bm{\eta}\left(\bm{p},\alpha\right)\right) \\
		&=\mathop{\arg\max}_{\left[\bm{p},\alpha\right]} \quad \left(\hat{\bm{\eta}}-\bm{\eta}\left(\bm{p},\alpha\right)\right)^T \bm{\Sigma}^{-1} \left(\hat{\bm{\eta}}-\bm{\eta}\left(\bm{p},\alpha\right)\right) ,
	\end{aligned}
\end{equation}
where $\bm{\eta}\left(\bm{p},\alpha\right)$ is the function characterizing the relationship between $\bm{\eta}$ and $\left[\bm{p},\alpha\right]$. In practice, $\bm{\eta}$ could be obtained via some compressive sensing techniques due to the sparsity of mmWave channels.

The goal of the multiple-RIS-aided positioning system is to minimize the expected estimation error of the MU's position and orientation, which is measured as 
\begin{align}
	&\rm{var}\left(\hat{\pmb{p}}\right)=\mathbb{E}\left[\left(\pmb{p}-\hat{\pmb{p}}\right)^T\left(\pmb{p}-\hat{\pmb{p}}\right)\right] , \\
	&\rm{var}\left(\alpha\right)=\mathbb{E}\left[\left(\alpha-\hat{\alpha}\right)^2\right] .
\end{align}

\section{Derivation of the Fundamental Bounds}
In this section, we derive the CRLBs of the estimation of the MU's position and the rotation angle, based on which we can obtain the position error bound (PEB) and rotation error bound (REB). To this end, we first derive the fisher information matrix (FIM) from the  channel parameters $\bm{\eta}$. 
The FIM of the aforementioned parameters is a $(6K+5)\times (6K+5)$ matrix $\bm{J_{\eta}}$ with the following expression
\begin{equation}
	\left[\bm{J}_{\bm{\eta}}\right]_{mn} = \frac{2 P_{TX}}{N_0 B}\sum_{n=1}^{N}\mathrm{Re}\left\{\frac{\partial \bm{\mu}^H [n]}{\partial \eta_m}\frac{\partial \bm{\mu} [n]}{\partial \eta_n}\right\} ,
\end{equation}
where $\bm{\mu}[n] =\bm{H}[n]\bm{F{\rm x}}[n]$ since it is an OFDM system.
The components in $\bm{J}_{\bm{\eta}}$ are specified as follows: 
\begin{align}
	&
	\frac{\partial \bm{\mu}[n]}{\partial \tau_{0}}=\gamma_0 h_0 \bm{\alpha}_{RX,0} \bm{\alpha}^H_{TX,0} \bm{F{\rm x}}[n] e^{j2\pi B \frac{n}{N}\tau_{0}} j2\pi B\frac{n}{N} , \\
	&\begin{aligned}
		\frac{\partial \bm{\mu}[n]}{\partial \tau_{k}}=&\gamma_k h_k \bm{\alpha}_{RX,k}\bm{\alpha}_{OUT,k}^{H}\bm{\Theta}_k \bm{\alpha}_{IN,k} \bm{\alpha}^H_{TX,k} \\
		& \bm{F{\rm x}}[n] e^{j2\pi B \frac{n}{N}\tau_{k}}  j2\pi B\frac{n}{N} ,
	\end{aligned} \\
	&\frac{\partial \bm{\mu}[n]}{\partial \theta_{TX,0}}=\gamma_0 h_0 \bm{\alpha}_{RX,0} \bm{\alpha}^H_{TX,0}\bm{D}^H_{TX,0}\bm{F{\rm x}}[n] ,\\
	&\frac{\partial \bm{\mu}[n]}{\partial \theta_{RX,0}}=\gamma_0 h_0 \bm{D}_{RX,0} \bm{\alpha}_{RX,0} \bm{\alpha}^H_{TX,0} \bm{F{\rm x}}[n]  ,\\
	& \begin{aligned}
		\frac{\partial \bm{\mu}[n]}{\partial \theta_{RX,k}}=&\gamma_k h_k \bm{D}_{RX,k} \bm{\alpha}_{RX,k} \bm{\alpha}_{OUT,k}^{H} \bm{\Theta}_k \\
		&\bm{\alpha}_{IN,k} \bm{\alpha}^H_{TX,k} \bm{F{\rm x}}[n] , \\
	\end{aligned} \\
	& \begin{aligned}
		\frac{\partial \bm{\mu}[n]}{\partial \varphi_{out,k}^{a}}=&\gamma_k h_k \bm{\alpha}_{RX,k} \bm{\alpha}_{OUT,k}^{H} {\rm diag}\left(\bm{c}^{a}_{out,k}\right)^H \\
		&\bm{\Theta}_k \bm{\alpha}_{IN,k} \bm{\alpha}^H_{TX,k} \bm{F{\rm x}}[n] ,
	\end{aligned} \\
	& \begin{aligned}
		\frac{\partial \bm{\mu}[n]}{\partial \varphi_{out,k}^{e}}=&\gamma_k h_k \bm{\alpha}_{RX,k}\bm{\alpha}_{OUT,k}^{H} {\rm diag}\left(\bm{c}^{e}_{out,k}\right)^H \\ 
		& \bm{\Theta}_k \bm{\alpha}_{IN,k} \bm{\alpha}^H_{TX,k} \bm{F{\rm x}}[n] ,
	\end{aligned} \\
	& \frac{\partial \bm{\mu}[n]}{\partial h_{R,0}}=\gamma_0 \bm{\alpha}_{RX,0} \bm{\alpha}^H_{TX,0} \bm{F{\rm x}}[n] e^{j2\pi B \frac{n}{N}\tau_{0}} ,\\
	& \frac{\partial \bm{\mu}[n]}{\partial h_{I,0}}=j\gamma_0 \bm{\alpha}_{RX,0} \bm{\alpha}^H_{TX,0} \bm{F{\rm x}}[n] e^{j2\pi B \frac{n}{N}\tau_{0}}  ,\\
	& \begin{aligned}
		\frac{\partial \bm{\mu}[n]}{\partial h_{R,k}}=&\gamma_k \bm{\alpha}_{RX,k}\bm{\alpha}_{OUT,k}^{H}\bm{\Theta}_k \bm{\alpha}_{IN,k} \bm{\alpha}^H_{TX,k} \\
		& \bm{F{\rm x}}[n] e^{j2\pi B \frac{n}{N}\tau_{k}} ,
	\end{aligned} \\
	& \begin{aligned}
		\frac{\partial \bm{\mu}[n]}{\partial h_{R,k}}=&j\gamma_k \bm{\alpha}_{RX,k}\bm{\alpha}_{OUT,k}^{H}\bm{\Theta}_k \bm{\alpha}_{IN,k} \bm{\alpha}^H_{TX,k} \\
		& \bm{F{\rm x}}[n] e^{j2\pi B \frac{n}{N}\tau_{k}} ,
	\end{aligned} 
\end{align}
where
\begin{equation}
\begin{aligned}
	& \bm{\alpha}_{TX,k}=\bm{\alpha}_{TX}\left(\theta_{TX,k}\right) , \quad \bm{\alpha}_{RX,k}=\bm{\alpha}_{RX}\left(\theta_{TX,k}\right),\\
	& \bm{\alpha}_{IN,k}=\bm{\alpha}_{RIS,IN}\left(\varphi_{in,k}^{a},\varphi_{in,k}^{e}\right)  ,\\
	& \bm{\alpha}_{OUT,k}=\bm{\alpha}_{RIS,OUT}\left(\varphi_{out,k}^{a},\varphi_{out,k}^{e}\right) ,\\
	& \!\!\!\!\!
	\bm{D}_{TX,k}(\theta)=j\frac{2\pi}{\lambda}d\cos(\theta_{TX,k}){\rm diag}\left(0,1,\cdots,N_t-1 \right) ,\\
	& \!\!\!\!\!
	\bm{D}_{RX,k}(\theta)=j\frac{2\pi}{\lambda}d\cos(\theta_{RX,k}){\rm diag}\left(0,1,\cdots,N_r-1 \right) ,\\
	& \!\!\!\!\!\!\!\!
	\left[\bm{c}_{out,k}^{a}\right]_{a+(b-1)L}=j\frac{2\pi}{\lambda}(a-1)d\cos(\varphi_{out,k}^{a})\sin(\varphi_{out,k}^{e}) , \\
	& \!\!\!\!\!\!\!\!
	\begin{aligned}
		\left[\bm{c}_{out,k}^{e}\right]_{a+(b-1)L}=&j\frac{2\pi}{\lambda}d[(a-1)\sin\left(\varphi_{out,k}^{a}\right) \\ &\cos\left(\varphi_{out,k}^{e}\right)-(b-1)\sin\left(\varphi_{out,k}^{e}\right) ] .
	\end{aligned}
\end{aligned}
\end{equation}

Next, the bijective matrix $\bm{T}$ is given by
\begin{equation}
	\!\!\!\!\!\!\!\!\!\!
	\bm{T}\!=\!\begin{bmatrix}
		\frac{\partial \bm{\tau}}{\partial p_x} &  \frac{\partial \theta_{TX,0}}{\partial p_x} &  \frac{\partial \bm{\theta}_{RX}}{\partial p_x} 
		&   \frac{\partial \bm{\varphi}_{out}^{a}}{\partial p_x} & \frac{\partial \bm{\varphi}_{out}^{e}}{\partial p_x}
		& \frac{\partial \bm{h}_{R}}{\partial p_x} &  \frac{\partial \bm{h}_{I}}{\partial p_x} \\
		\frac{\partial \bm{\tau}}{\partial p_y} &  \frac{\partial \theta_{TX,0}}{\partial p_y} &  \frac{\partial \bm{\theta}_{RX}}{\partial p_y} 
		& \frac{\partial \bm{\varphi}_{out}^{a}}{\partial p_y} &  \frac{\partial \bm{\varphi}_{out}^{e}}{\partial p_y} 
		& \frac{\partial \bm{h}_{R}}{\partial p_y} &  \frac{\partial \bm{h}_{I}}{\partial p_y} \\ 
		\frac{\partial \bm{\tau}}{\partial \alpha} & \frac{\partial \theta_{TX,0}}{\partial \alpha}  & \frac{\partial \bm{\theta}_{RX}}{\partial \alpha} 
		& \frac{\partial \bm{\varphi}_{out}^{a}}{\partial \alpha} &  \frac{\partial \bm{\varphi}_{out}^{e}}{\partial \alpha} 
		& \frac{\partial \bm{h}_{R}}{\partial \alpha} &  \frac{\partial \bm{h}_{I}}{\partial \alpha} 
	\end{bmatrix}.
\end{equation}
By utilizing the following geometric relationships among the BS, RISs and the MU shown in Fig. \ref{img-1d3r}
\begin{align}
	&\tau_0=\left\| \bm{q}-\bm{p} \right\| /c , \quad \tau_{k}=\left\| \bm{q}-\bm{s}_k \right\| /c + \left\| \bm{p}-\bm{s}_k \right\| /c , \\
	&\theta_{TX,0}=\arcsin \left( \left( p_x-q_x \right)/\left\| \bm{q}-\bm{p} \right\| _2 \right) , \\
	&\varphi_{out,k}^{a}=\arcsin\left( \frac{p_y-s_{ky}}{\sqrt{\left( p_x-s_{kx} \right) ^2+\left( p_y-s_{ky} \right) ^2}} \right) , \\
	&\varphi _{out,k}^{e}=\arccos \left( s_{kz}/\left\| \bm{p}-\bm{s}_k \right\| _2 \right) , \\
	&\theta_{RX,0}=\arcsin \left( \frac{\left( p_x-q_x \right) \cos \alpha -\left( p_y-q_y \right) \sin \alpha}{\left\| \bm{q}-\bm{p} \right\| _2} \right) , \\
	& \!\!\!\!\!\!
	\theta_{RX,k}=\arcsin \left( \frac{\left( p_x-s_{kx} \right) \cos \alpha -\left( p_y-s_{ky} \right) \sin \alpha}{\left\| \bm{p}-\bm{s}_k \right\| _2} \right) ,
\end{align}
the non-zero entries of the bijective transformation matrix $\bm{T}$ are given by 
\begin{align}
	&\frac{\partial \tau_0}{\partial p_x}=\frac{1}{c}\frac{p_x-q_x}{\left\| \bm{q}-\bm{p} \right\| _2} , \quad
	\frac{\partial \tau_0}{\partial p_y}=\frac{1}{c}\frac{p_y-q_y}{\left\| \bm{q}-\bm{p} \right\| _2} , \\
	&\frac{\partial \tau_k}{\partial p_x}=\frac{1}{c}\frac{p_x-s_{kx}}{\left\| \bm{p}-\bm{s}_k \right\| _2} , \quad 
	\frac{\partial \tau_k}{\partial p_y}=\frac{1}{c}\frac{p_y-s_{ky}}{\left\| \bm{p}-\bm{s}_k \right\| _2} , \\
	&\frac{\partial \theta _{TX,0}}{\partial p_x}=\frac{\sqrt{\left\| p-q \right\| _{2}^{2}-\left( p_x-q_x \right) ^2}}{\left\| \bm{q}-\bm{p} \right\| _{2}^{2}} , \\
	&\frac{\partial \theta _{TX,0}}{\partial p_y}=\frac{\left( p_x-q_x \right) \left( p_y-q_y \right)}{\sqrt{\left\| \bm{q}-\bm{p} \right\| _{2}^{2}-\left( p_x-q_x \right) ^2}\cdot \left\| \bm{q}-\bm{p} \right\| _{2}^{2}} , \\
	&\frac{\partial \varphi _{out,k}^{a}}{\partial p_x}=-\frac{p_y-s_{ky}}{\left( p_x-s_{kx} \right) ^2+\left( p_y-s_{ky} \right) ^2} , \\
	&\frac{\partial \varphi _{out,k}^{a}}{\partial p_y}=\frac{p_x-s_{kx}}{\left( p_x-s_{kx} \right) ^2+\left( p_y-s_{ky} \right) ^2} , \\
	&\frac{\partial \varphi _{out,k}^{e}}{\partial p_x}=\frac{s_{kz}\left( p_x-s_{kx} \right)}{\left\| p-s_k \right\| _{2}^{2}\sqrt{\left\| p-s_k \right\| _{2}^{2}-\beta _{IRS}^{2}}} , \\
	&\frac{\partial \varphi _{out,k}^{e}}{\partial p_y}=\frac{s_{kz}\left( p_y-s_{ky} \right)}{\left\| p-s_k \right\| _{2}^{2}\sqrt{\left\| p-s_k \right\| _{2}^{2}-\beta _{IRS}^{2}}} , \\
	&\frac{\partial \theta _{RX,0}}{\partial p_x}\!=\!\frac{\cos \alpha -\frac{\left( p_x-q_x \right) \left[ \left( p_x-q_x \right) \cos \alpha -\left( p_y-q_y \right) \sin \alpha \right]}{\left\| p-q \right\| _{2}^{2}}}{\sqrt{\left\| \bm{q}-\bm{p} \right\| _{2}^{2}-\left[ \left( p_x-q_x \right) \cos \alpha -\left( p_y-q_y \right) \sin \alpha \right] ^2}} , \\
	&\frac{\partial \theta _{RX,0}}{\partial p_y}\!=\!\frac{\sin \alpha +\frac{\left( p_y-q_y \right) \left[ \left( p_x-q_x \right) \cos \alpha -\left( p_y-q_y \right) \sin \alpha \right]}{\left\| p-q \right\| _{2}^{2}}}{\sqrt{\left\| \bm{q}-\bm{p} \right\| _{2}^{2}-\left[ \left( p_x-q_x \right) \cos \alpha -\left( p_y-q_y \right) \sin \alpha \right] ^2}} , \\
	&\frac{\partial \theta _{RX,0}}{\partial \alpha}\!=\!\frac{-\left( p_x-q_x \right) \sin \alpha -\left( p_y-q_y \right) \cos \alpha}{\sqrt{\left\| \bm{q}-\bm{p} \right\| _{2}^{2}-\left[ \left( p_x-q_x \right) \cos \alpha -\left( p_y-q_y \right) \sin \alpha \right] ^2}} , \\
	&\!\!\!\!\!\!
	\frac{\partial \theta _{RX,k}}{\partial p_x}\!=\!\frac{\cos \alpha -\frac{\left( p_x-s_{kx} \right) \left[ \left( p_x-s_{kx} \right) \cos \alpha -\left( p_y-s_{ky} \right) \sin \alpha \right]}{\left\| p-s_k \right\| _{2}^{2}}}{\sqrt{\left\| \bm{p}-\bm{s}_k \right\| _{2}^{2}-\left[ \left( p_x-s_{kx} \right) \cos \alpha -\left( p_y-s_{ky} \right) \sin \alpha \right] ^2}} , \\
	&\!\!\!\!\!\!
	\frac{\partial \theta _{RX,k}}{\partial p_y}\!=\!\frac{\sin \alpha +\frac{\left( p_y-s_{ky} \right) \left[ \left( p_x-s_{kx} \right) \cos \alpha -\left( p_y-s_{ky} \right) \sin \alpha \right]}{\left\| p-s_k \right\| _{2}^{2}}}{\sqrt{\left\| \bm{p}-\bm{s}_k \right\| _{2}^{2}-\left[ \left( p_x-s_{kx} \right) \cos \alpha -\left( p_y-s_{ky} \right) \sin \alpha \right] ^2}} , \\
	&\!\!\!\!\!\!
	\frac{\partial \theta _{RX,k}}{\partial \alpha}\!=\!\frac{-\left( p_x-s_{kx} \right) \sin \alpha -\left( p_y-s_{ky} \right) \cos \alpha}{\sqrt{\left\| \bm{p}-\bm{s}_k \right\| _{2}^{2}-\left[ \left( p_x-s_{kx} \right) \cos \alpha -\left( p_y-s_{ky} \right) \sin \alpha \right] ^2}} .
\end{align}

Finally, we obtain the FIM $\bm{J}$ in position domain through $\bm{J}_{\eta}$ and $\bm{T}$,
\begin{equation}
	\bm{J}=\bm{TJ_{\eta}T}^H .
\end{equation}
Thus, the PEB is derived as the root square of the trace of the first $2\times2$ sub-matrix of $\bm{J}$
\begin{equation}
	{\rm PEB}=\sqrt{\mathrm{tr}\left(\bm{J}_{1:2,1:2}^{-1}\right)} ,
\end{equation}
and the REB is given by the root square of the third diagonal entry of $\bm{J}$
\begin{equation}
	{\rm REB}=\sqrt{\bm{J}_{3,3}^{-1}} .
\end{equation}

\section{Phase Shift Optimization}
Based on the derived CRLB expression, it is observed that the phase shifts of the RIS have a great impact on the CRLB. In this paper, we aim to optimize the phase shifts of the RIS to minimize the sum PEB and REB, which is formally formulated as follows:
\begin{equation}
\begin{aligned}
	\min_{\bm{\Theta}_k} \quad & \rm{PEB+REB} \\
	\rm{s.t.} \quad & \bm{\Theta_{k}}=\delta \times \rm{diag}\left(e^{j\theta_{1}},e^{j\theta_2},\cdots,e^{j\theta_{L^2}}\right) ,\\
	& \theta_i \in [0,2\pi), \quad i=1,2,\cdots,L^2 .
\end{aligned}
\label{op problem}
\end{equation}
However, the objective function in Problem (\ref{op problem}) is too complex and the conventional gradient method is not applicable. To address this issue, the heuristic method such as PSO is adopted, which only needs to evaluate the objective function value in each iteration rather than calculating the fisrt-order derivative of the original objective function that entails high computational complexity.

\section{Simulation Results}
Consider a positioning system with one direct path and three reflecting paths separately going through three RISs. The BS is located at $\rm{(0\, m, 0\, m, 40\, m)}$ and MU is located at $\rm{(90\, m,30\, m,0\, m)}$. The locations of the RISs are $\rm{(60\, m, 45\, m, 15\, m)}$,$\rm{(50\, m, 50\, m, 5\, m)}$ and $\rm{(40\, m, 20\, m, 10\, m)}$. The number of reflecting elements at the RIS is set as $L^2=16^2=256$. The numbers of transmitter and receiver antennas are $N_t=32$ and $N_r=8$ respectively. The carrier frequency is $f_c=4.9$ GHz and the number of subcarriers is $N=128$. The noise power spectrum density is $N_0=-174$ dBm/Hz, bandwidth is $B=20\,$MHz, the path loss exponent of the direct channel is $\alpha_0=3.7$, the path loss exponent of the $k$-th RIS is $\alpha_k=2.2$ and shadow fading parameters of direct path and reflecting path are respectively $\sigma_{SF,0}=4$ and $\sigma_{SF,k}=7$.

\subsection{The Impact of the Number of RISs}
In Fig. \ref{PREB}, we compare the performance of our optimized phase shifts of the RIS with the case when the phase shifts are randomly set. It is observed from Fig. \ref{PREB} that the optimized phase shifts can achieve much better performance than the randomly generated phase shifts in terms of both PEB and REB. In addition, Fig. \ref{PREB} shows that PEB and REB gradually decreases with the number of RISs, which demonstrates the advantages of using multiple RISs in positioning. 

\subsection{The Impact of the Phase Shifts of RIS}
\begin{figure}[htbp]
	\centering
	\subfigure[PEB]{
		\centering
		\includegraphics[width=1 \linewidth]{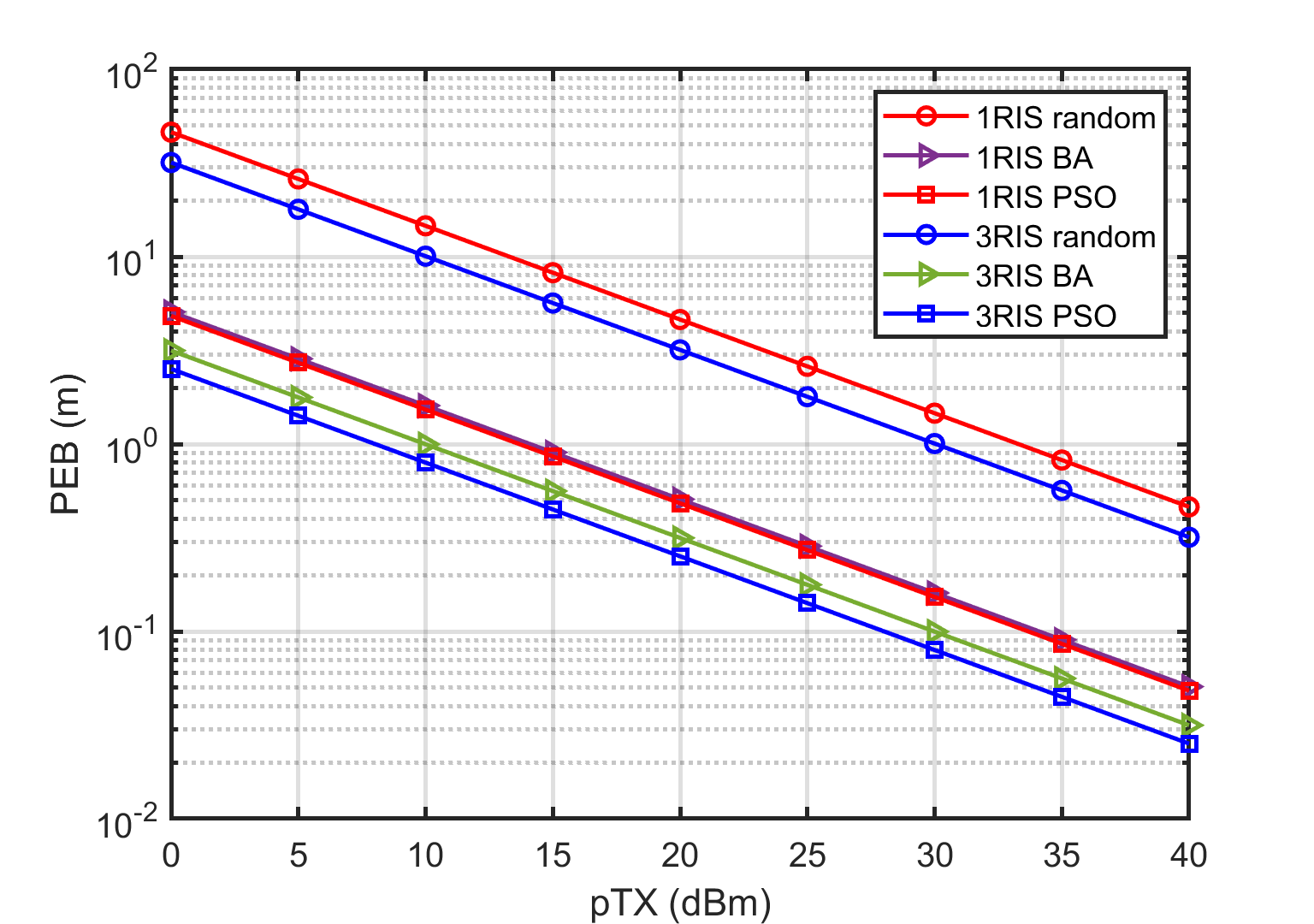}
	}
	\subfigure[REB]{
		\centering
		\includegraphics[width=1 \linewidth]{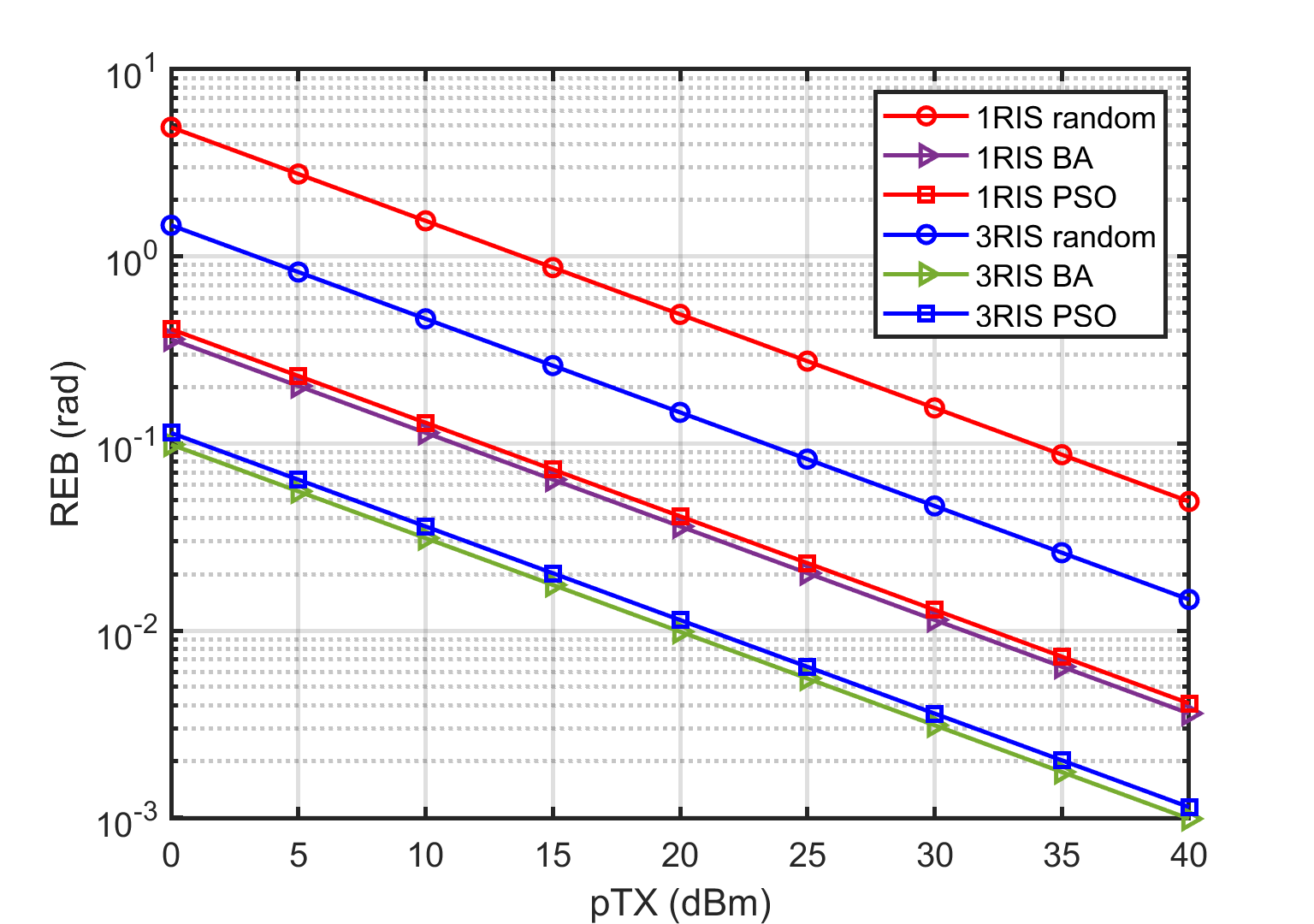}
	}
	\caption{The PEB and REB with different number of paths and different kinds of phase shifts}
	\label{PREB}
\end{figure}

Concerning the optimization of the phase shifts, one method is to set the phase shifts that align with the channel vectors as in \cite{JH2020,2021RISwangrui}. Specifically, the phase shifts of the RIS are set as follows: 
\begin{equation}
	\begin{aligned}
		\bm{\Theta}_{k}^{\prime}=\delta \times {\rm diag}&\left\{\bm{\alpha}_{RIS,IN}^{*}\left(\varphi_{in,k}^{a},\varphi_{in,k}^{e}\right) 
		\right. \\ \phantom{=\;\;} &  \left. 
		\odot \bm{\alpha}_{RIS,OUT}\left(\varphi_{out,k}^{a},\varphi_{out,k}^{e}\right)  \right\} .
	\end{aligned} 
\end{equation} 
However, this approach only maximizes part of the elements in the FIM, and cannot ensure that each entry of the FIM is maximized, especially for those elements containing $\frac{\partial \bm{\mu}[n]}{\varphi_{out,k}^{a}}$ and $\frac{\partial \bm{\mu}[n]}{\varphi_{out,k}^{e}}$. On the other hand, the PSO algorithm aims to search the globally optimal solution.
From Fig. \ref{PREB}, we find that when the system has  only one RIS, the performance of the PSO-optimized phase shifts are close to that of the beam-aligned (BA) phase shifts. Furthermore, in the case of multiple RISs, although the REBs of two schemes are still close, the PSO-optimized phase shifts performs much better than the beam-aligned phase shifts  in terms of the PEB performance.

\subsection{The Impact of the Size of RIS}
\begin{figure}[htbp]
	\centering \subfigure[PEB]{ \centering \includegraphics[width=0.45\linewidth]{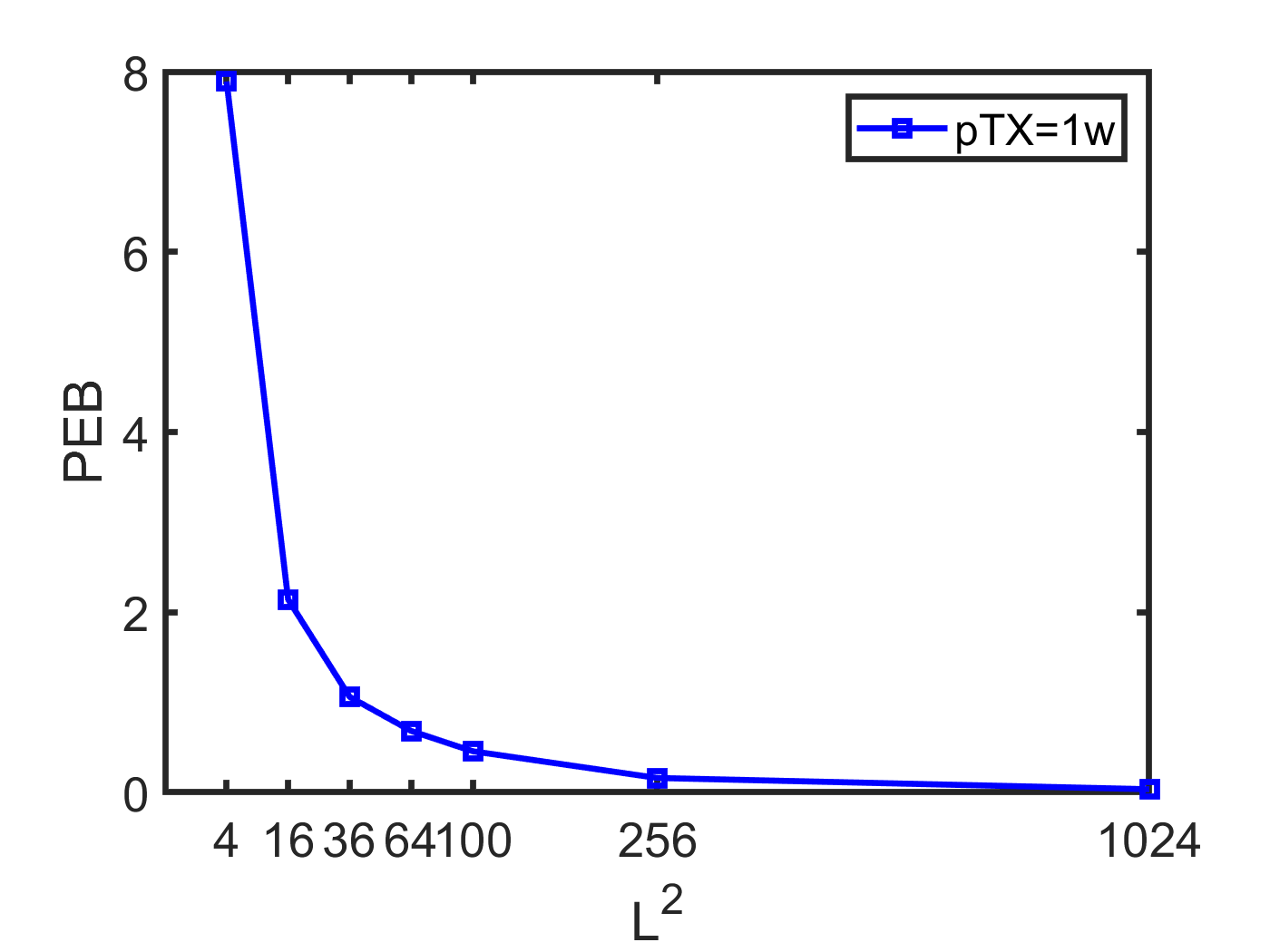}
	} \subfigure[REB]{ \centering \includegraphics[width=0.45\linewidth]{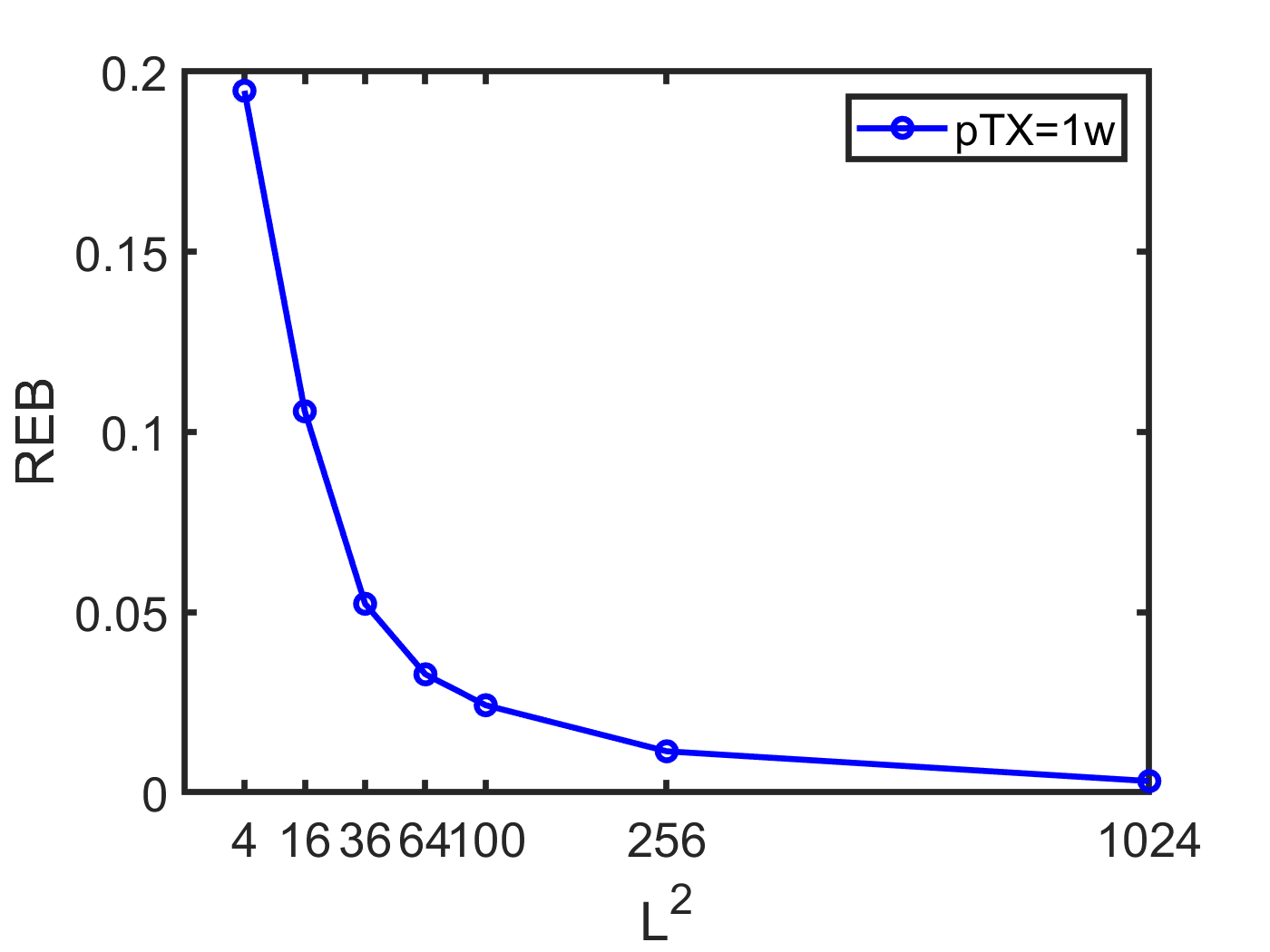}
	} \caption{The PEB and REB with different size of RISs}
	\label{size}
\end{figure}

Fig. \ref{size} illustrates the impact of the size of RIS on the positioning performance. It is readily seen that more reflecting elements could contribute to a higher resolution, which results in a higher quality of positioning. It is interesting to find that the improvement is significant when the size is small and will reduce as the size increases. 

\section{Conclusion}
In this paper, the Cram$\rm{\acute{e}}$r-Rao bound of a multiple-RIS-aided mmWave MIMO positioning system has been derived. Through the Cram$\rm{\acute{e}}$r-Rao bound of location and rotation angle estimation, termed as PEB and REB respectively, we have investigated the impact of the number, the sizes, and the phase shifts of RISs on positioning performance. Simulation results have shown the improvement in PEB and REB by adopting multiple RISs, indicating the great potential of RISs to help achieve high positioning accuracy.

\bibliographystyle{IEEEtran}
\bibliography{myre}
\end{document}